\documentclass{svproc}

\usepackage{url}
\usepackage[english]{babel}
\usepackage{amsmath}
\usepackage{cite}
\usepackage{amsfonts}
\usepackage{amssymb}
\usepackage{graphicx}
\usepackage{gensymb}
\usepackage{subfig}
\usepackage{float}
\usepackage{wrapfig,lipsum,booktabs}
\usepackage[colorlinks=true, allcolors=blue]{hyperref}
\usepackage{stackrel} 
\usepackage{polynom} 
\polyset{style=A, div=:, vars=x}
\usepackage{xcolor}
\usepackage{authblk}
\usepackage[normalem]{ulem}
\usepackage{multirow}
\usepackage{tikz, todonotes}
\usepackage{colortbl}
\usetikzlibrary{shapes,shadows,arrows}

\newcommand{\defeq}{:=}

\begin{document}
\mainmatter 
\title{Mathematical model for fluoride-removal filters}
\titlerunning{Modelling fluoride-removal filtration} 

\author{Lucy C. Auton\inst{1}\footnote{Corresponding author: lauton@crm.cat} \and Marc Martínez I Àvila\inst{2}\and 
Shanmuk S. Ravuru\inst{3} \and Sirshendu De\inst{3} 
\and Tim G. Myers\inst{1} \and
Abel Valverde\inst{4}}
\authorrunning{ Auton \& Valverde et al.} 

\tocauthor{Lucy C. Auton, Marc Martínez I. Àvila,
Shanmuk S. Ravuru, Sirshendu De, Tim Myers and Abel Valverde}

\institute{Centre de Recerca Matem\`{a}tica, Campus de Bellaterra, Barcelona, Spain \\
\and
Dept. Math., Universitat Autònoma de Barcelona, 
Spain 
\and 
Dept. Chem. Eng., Indian Institute of Technology Kharagpur, India \and
Dept. Chem. Eng, Universitat Polit\`{e}cnica de Catalunya, Barcelona, Spain}

\maketitle 

\begin{abstract}
We develop a model that captures the dominant chemical mechanisms involved in the removal of fluoride from water by a novel adsorbent comprising mineral rich carbon (MRC) and chemically treated mineral rich carbon (TMRC). Working with experimental data, we validate the model for both MRC and TMRC based on the underlying chemical reactions. 
The model we derive from the chemical composition
of TMRC and MRC shows good agreement with experimental results.

\end{abstract}

\section{Introduction}

Small amounts of fluoride (F$^-$) are necessary for healthy teeth and bone growth; many toothpastes contain fluoride to aid in dental hygiene \cite{kaminsky1990fluoride}. However excess amounts of fluoride can lead to dental and skeletal fluorosis amongst other afflictions, including cancer, gastro-intestinal distress and brain damage \cite{kaminsky1990fluoride, bharti2017fluoride}. Fluoride occurs naturally in certain rocks and soils, from which it leaches into ground water; additionally, fluoride can be found in industrial run-off and 
 aerosols which affects the ecosystem \cite{ozsvath2009fluoride}. For example, in India it is estimated that 62 million people are consuming water with more than 1.5mg/l, which is the maximum concentration recommended by the WHO \cite{world2019preventing}. 

Here, we consider two different fluoride adsorbents: mineral rich carbon (MRC) and chemically treated mineral rich carbon (TMRC). These adsorbents have proven highly effective in removing fluoride from water. Chatterjee et al. \cite{chatterjee2018defluoridation} 
find that TMRC (therein referred to as chemically treated carbonised bone meal or CTBM) has a capacity of 150mg/g while the next highest adsorption capacity of alternative bio-based adsorbents is aluminum-treated activated bamboo charcoal, with an adsorption capacity of 21.1mg/g \cite{wendimu2017aluminium}. MRC (referred to as carbonised bone meal or CBM in Chatterjee et al. \cite{chatterjee2018novel}) has an adsorption capacity of 14mg/g. Bhatnagar et al. \cite{bhatnagar2011fluoride} consider 102 adsorbents for F$^-$ and find only four which have a higher adsorption capacity than TMRC (Nanomagnesia, CaO nanoparticles, Calcined Mg-Al-CO3 and Fe–Al–Ce trimetal oxide). 

Using the underlying chemical structure of MRC and TMRC in combination with new batch experiments, we develop individual models for the removal of fluoride using MRC and TMRC. The combined MRC-TMRC model has good agreement with the breakthrough curves provided by the column experiments, while for physically reasonable parameters, Langmuir's model shows poor agreement. 

\section{Model presentation and validation}
We consider the filtration of water contaminated with the passive solute fluoride, (F$^-$), by the novel adsorbents MRC \cite{chatterjee2018novel}, and TMRC \cite{chatterjee2018defluoridation}. MRC is carbonised mammalian or avian bone meal, while TMRC is a derivative of MRC which has been chemically treated to improve its fluoride adsorption capacity. TMRC is made by grinding down MRC and coating it in Aluminium Hydroxide $\big(\text{Al(OH)}_3\big)$. 
This treatment process makes TMRC more expensive than MRC, since the cost of aluminium significantly exceeds the cost of bone meal. Further, TMRC has an average grain size of 0.1--0.3mm which is notably smaller than MRC, which has an average grain size of 0.4--0.6mm \cite{chatterjee2018defluoridation}. We consider two different filters: batch filters, which are experimental tools (Figure~\ref{Batch diagram}, left) and column filters, which are used in practice (Figure~\ref{Batch diagram}, right). 

\subsection{Materials and methods}
\begin{figure}[htb]
 \centering 
 \includegraphics[width=.95\textwidth]{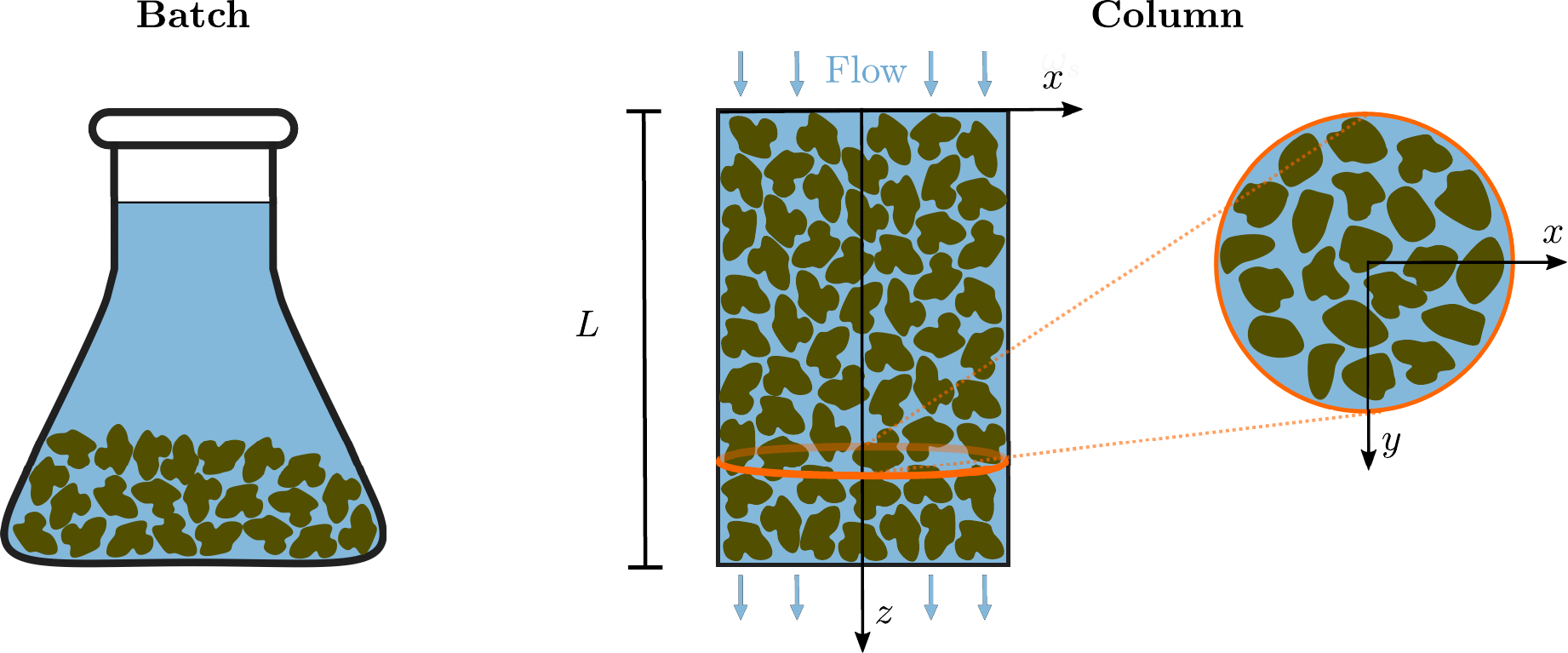}
 \caption{ 
 Schematic of batch experiments (left) and column experiments (right). 
 \textbf{\textit{Left}}: A beaker contains a small amount of adsorbent (MRC or TMRC) surrounded by contaminated fluid; this forms a closed system with no net flow. 
 \textbf{\textit{Right}}: Contaminated fluid flows through a cylinder evenly packed with a mixture of 40:1 (MRC:TMRC).
 }
 \label{Batch diagram}
\end{figure}

The materials MRC and TMRC are created following the methodology detailed in Chatterjee et al. \cite{chatterjee2018novel,chatterjee2018defluoridation} and a full analysis of their properties is given therein. We consider two different experimental setups: batch (Figure \ref{Batch diagram}, left) and column (or fixed-bed) adsorption experiments (Figure \ref{Batch diagram}, right). 
The batch experimental setup enables us to determine the intrinsic properties of the fluid--adsorbate--adsorbent system, while the column filter is used in both prototype and commercialised filters.

A batch experiment is one in which fluid contaminated with a known concentration of fluoride, $c_\mathrm{F}$, is added to a beaker (Figure \ref{Batch diagram}, left), which contains a small amount of adsorbent (7g/l). The beaker is then placed in a shaker. Batch experiments are classified into two types: isotherm studies and kinetic studies. We use experimental data for both the kinetic and isotherm studies for MRC and TMRC, separately.
For the isotherm study, the beaker remains in the shaker until equilibrium is reached, at which point the final concentration is measured; this is repeated for different initial concentrations of fluoride, $c_\mathrm{F}^\mathrm{i}$, 
This produces an isotherm curve which gives the equilibrium value of the adsorbed mass of contaminant, per unit mass of adsorbent, $q^\mathrm{e}$, against the corresponding equilibrium concentration of fluoride, $c_\mathrm{F}^\mathrm{e}$. 
For the kinetic study, $c_\mathrm{F}$, is measured at various times by removing a small amount of the fluid; one experiment provides data for an entire kinetic curve. 

The column filters comprise a glass cylinder of internal diameter 4.4 cm and length 25 cm packed with a homogeneous mixture of MRC and TMRC at a ratio of 40:1 (MRC:TMRC) by mass (Figure \ref{Batch diagram}, right). The MRC-TMRC mixture is evenly packed to a height of 10cm over a base of sand; fluid flows through the filter by gravity. The operating pH is 7 and the ambient temperature is 300$\pm$3.0 K. 
Samples are collected at the outlet at regular intervals and the outlet concentration, $c_\mathrm{F}^\mathrm{out}$, is determined using an ion-selective electrode. 
\subsection{Batch modelling}

Batch experiments involve a closed system, as such the only change in $c_\mathrm{F}$ is via adsorption to the surface of the MRC or TMRC giving
\begin{subequations}
\begin{equation}
 \label{equation system Langmuir}
 \frac{\partial c_\mathrm{F}}{\partial t} =-\frac{\rho^\mathrm{B}_\ast}{\phi_\ast}\frac{\partial q_\star}{\partial t}, 
 \end{equation}
 where $t$ is time, $q_\star$ is the adsorbed mass of contaminant, per unit mass of adsorbent according to model $\star$ where $\star\in\{\text{M,T,C,L}\}$ represents the chemically based models for MRC, TMRC, combined MRC-TMRC mixture and Langmuir's model, respectively, and $\rho^\mathrm{B}_\ast$ is the bulk density which is defined to be the initial mass of adsorbent $\ast\in\{\text{M,T,C}\}$, denoting MRC, TMRC and the MRC-TMRC mixture respectively, divided by the total volume of the filter, $|\Omega|$, and where we define the porosity $\phi_\ast$ to be the fluid fraction in $\Omega$ for the adsorbent $\ast$. 

 \subsubsection{MRC} 
 Contaminant removal is modelled via $\partial q_\star/\partial t$; the standard approach, derived for physisorption or a one-to-one contaminant-to-adsorbate chemisorption, is Langmuir's equation
\begin{equation}
 \frac{\partial q_\mathrm{L}}{\partial t} =k_\mathrm{L}^\mathrm{a} c_\mathrm{F}(q_\mathrm{L}^\mathrm{m}-q_\mathrm{L})-k_\mathrm{L}^\mathrm{d} q_\mathrm{L}, \quad 
 \label{eq: Langmuir}
\end{equation}
\label{Lang_MRC}
\end{subequations}
 \hspace{-2.5mm} where $q_\mathrm{L}^\mathrm{m}$ 
 is the maximum amount absorbed
 (according to Langmuir's model), and where 
 $k_\mathrm{L}^\mathrm{a}$ and $k_\mathrm{L}^\mathrm{d}$ are known as the forwards and backwards reaction rates, respectively.
Note that Langmuir's model is defined by Equations~(\ref{Lang_MRC}), taking $\star = \mathrm{L}$. 
\begin{table}[b]
 \centering \hspace{-2mm}\footnotesize{
 \begin{tabular}{|c|c|c|c|c|c|c|c|c|}
 \hline 
 \multicolumn{9}{|c|}{\textbf{MRC isotherm and kinetic parameters (3 s.f.)}} \\ 
 \hline
 \multicolumn{3}{|c|}{\textbf{Extracted (kinetic)}} & \multicolumn{3}{|c|}{\textbf{Optimised: Langmuir}} & \multicolumn{3}{|c|}{\textbf{Optimised: CB-MRC}} \\
 \hline
 Param. & Value & Units & Param. & Value & Units & Param. & Value & Units \\
 \hline
 $c_\mathrm{F}^\mathrm{i}$ & 5.26$\times 10^{-4}$ & mol/l & $ \cellcolor[rgb]{0.7922,0.8471,0.8706}k_\mathrm{L}^\mathrm{a}/\kappa_\mathrm{L}^\mathrm{d}$& \cellcolor[rgb]{0.7922,0.8471, 0.8706}295 & \cellcolor[rgb]{0.7922,0.8471, 0.8706}l/mol & $ \cellcolor[rgb]{0.7922,0.8471, 0.8706}k_1^\mathrm{a}/k_1^\mathrm{d}$ & \cellcolor[rgb]{0.7922,0.8471, 0.8706}6.54 & \cellcolor[rgb]{0.7922,0.8471, 0.8706}-- \\
 \hline
 $c_\mathrm{OH}^\mathrm{i}$ & $1.00\times10^{-8}$ & mol/l & \cellcolor[rgb]{0.7922,0.8471, 0.8706}$q_\mathrm{L}^\mathrm{m}$ & \cellcolor[rgb]{0.7922,0.8471, 0.8706}$7.27\times10^{-4}$ & \cellcolor[rgb]{0.7922,0.8471, 0.8706}mol/g & \cellcolor[rgb]{0.7922,0.8471, 0.8706}$k_2^\mathrm{a}/\kappa_2^\mathrm{d}$ &\cellcolor[rgb]{0.7922,0.8471, 0.8706} 99.7 & \cellcolor[rgb]{0.7922,0.8471, 0.8706}l/mol\\ 
 \hline
 $c_\mathrm{F}^\mathrm{e}$ & 1.05$\times 10^{-5}$ & mol/l & \cellcolor[rgb]{1,1, 0.8}$k_\mathrm{L}^\mathrm{a}$ & \cellcolor[rgb]{1,1, 0.8}0.126 & \cellcolor[rgb]{1,1, 0.8}l/(mol·s) & \cellcolor[rgb]{0.7922,0.8471, 0.8706}$ q_2^\mathrm{m}/q_\mathrm{M}^\mathrm{m}$ & \cellcolor[rgb]{0.7922,0.8471, 0.8706}0.790 & \cellcolor[rgb]{0.7922,0.8471, 0.8706}\% \\
 \hline
 \multicolumn{3}{|c|}{} & \multicolumn{3}{|c|}{} & \cellcolor[rgb]{0.7922,0.8471, 0.8706}$ q_\mathrm{M}^\mathrm{m}$ & \cellcolor[rgb]{0.7922,0.8471, 0.8706}8.45$\times 10^{-4}$ & \cellcolor[rgb]{0.7922,0.8471, 0.8706}$\text{mol}/\text{g}$\\
 \hline
 \multicolumn{3}{|c|}{} & \multicolumn{3}{|c|}{} &\cellcolor[rgb]{1,1, 0.8} $k_1^\mathrm{a}$ & \cellcolor[rgb]{1,1, 0.8}1.33 & \cellcolor[rgb]{1,1, 0.8}l/(mol·s) \\
 \hline
 \multicolumn{3}{|c|}{} & \multicolumn{3}{|c|}{} & \cellcolor[rgb]{1,1, 0.8}$k_2^\mathrm{a}$ & \cellcolor[rgb]{1,1, 0.8}2.08 & \cellcolor[rgb]{1,1, 0.8}l/(mol·s)\\
 \hline
 \end{tabular}
 }
 \normalsize
 \caption{Parameters for Langmuir removal (Eq.~\ref{Lang_MRC}) and for the chemically based model for MRC (CB-MRC, Eqs.~\ref{MRC eq system}), where $q_\mathrm{M}^\mathrm{m}\defeq q_1^\mathrm{m}+q_2^\mathrm{m}$. Those shaded in grey were determined via fitting the isotherm (Figure~\ref{MRC Lang}, left) and those shaded in pale yellow were subsequently fitted using the kinetic study (Figure~\ref{MRC Lang}, right).}
 \label{MRC new model table}
\end{table}

\begin{figure}[htb]
 \centering
 \includegraphics[width=.95\textwidth]{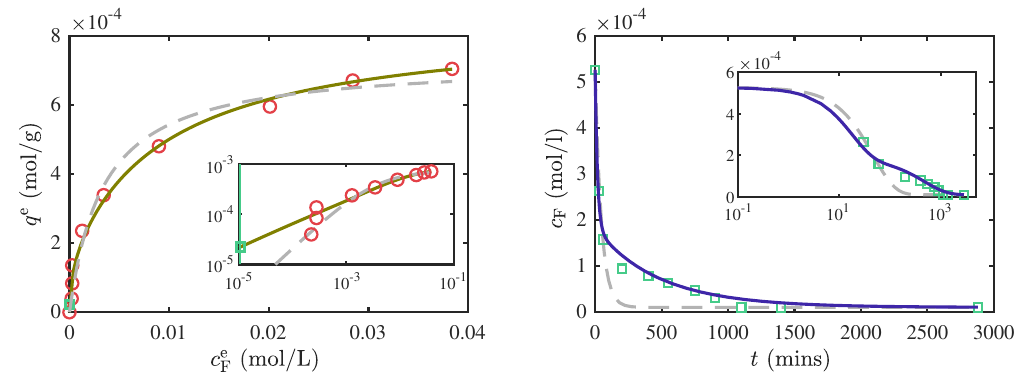}
 \caption{
 Isotherm (left) and kinetic study (right) for MRC fitted with both the CB-MRC model (Eqs.~\ref{MRC eq system}, solid coloured curve) and Langmuir (Eqs.~\ref{Lang_MRC} with $\star =$L, grey dashed curve). 
 \textit{\textbf{Left}}: All experimental data points (red circles) are the equilibrium values of $q$ against $c_\mathrm{F}$ for individual experiments. 
 We highlight the the $c^\mathrm{e}_\mathrm{F}$ experiment for which we have a kinetic study (green vertical line), and the corresponding value of $q^\mathrm{e}$ as predicted by the CB-MRC model (green square). 
 To examine the fit more closely for small $c_\text{F}^\text{e}$, we also include a log-log plot of $q^\text{e}$ against $c_\text{F}^\text{e}$ (inset). 
 \textit{\textbf{Right}}:~All experimental data points (green squares) are from the same experiment. 
 We include a semi-log plot of $c_\text{F}$ against $t$ (inset). 
 \label{MRC Lang}
 }
\end{figure}

 We use the isotherm to fit $k_\mathrm{L}^\mathrm{a}/\kappa_\mathrm{L}^\mathrm{d}$ and $q_\mathrm{L}^\mathrm{m}$; subsequently we use the kinetic study to fit $k_\mathrm{L}^\mathrm{a}$. All fitting is done in \verb|MATLAB|$^\text{\copyright}$, using the \verb|GlobalSearch| function. 
Table~\ref{MRC new model table} (centre columns), shows these parameters. 
Figure~\ref{MRC Lang} (left, grey dashed lines) shows that Langmuir's isotherm fits passably with the experimental data; however this model fails to capture the piecewise behaviour in the kinetic study (Figure~\ref{MRC Lang}, right, grey dashed lines). Thus, we infer that the adsorbent MRC does not adsorb $F^-$ via a simple one-to-one chemisorption reaction and hence must consider the underlying chemistry further. A MRC molecule consists of ten Calcium (Ca$^{2+}$) ions, six Phosphate (PO$_4^{3-}$) molecules and two Hydroxide ($\text{OH}^-$) molecules. Appealing to Sundaram et al. \cite{sundaram2008defluoridation} and Balasooriya et al. \cite{balasooriya2022applications}, we assert that the dominant chemical reactions occurring in MRC are 
\begin{subequations}

 \begin{align}
 \text{POH}+\text{F}^-&\stackrel[k_1^\mathrm{d}]{k_1^\mathrm{a}}{\rightleftharpoons}\text{PF}+\text{OH}^-, \label{MRC reactions Chemical}\\
 \text{PCa}+\text{F}^-&\stackrel[\kappa_2^\mathrm{d}]{k_2^\mathrm{a}}{\rightleftharpoons}\text{PCa}\leftrightsquigarrow \text{F},\label{MRC reactions Physical}
 \end{align}
 \label{MRC reactions}
\end{subequations}
 \hspace{-1.6mm}where P is the Hydroxyapatite lattice to which the OH$^{-}$ is bonded and where $k_1^\mathrm{a}$, $k_1^\mathrm{d}$ are the 
forwards and backwards reaction rates, respectively, of Equation (\ref{MRC reactions Chemical}) and $k_2^\mathrm{a}$, $\kappa_2^\mathrm{d}$ are the 
forwards and backwards reaction rates, respectively, of Equation (\ref{MRC reactions Physical}). Note that the units of $\kappa_2^\mathrm{d}$ are distinct from $k_1^\mathrm{d}$, motivating our notation. Equation (\ref{MRC reactions Chemical}) is a chemical reaction describing ion exchange between F$^-$ and OH$^-$, while Equation (\ref{MRC reactions Physical}) is a physisorption reaction, with $\leftrightsquigarrow$ denoting an electrostatic bond.
We express Equations~(\ref{MRC reactions}) as a system of differential equations 
\begin{subequations}
 \begin{align}
 &\frac{\partial c_\mathrm{F}}{\partial t} = -\frac{\rho^\mathrm{B}_\mathrm{M}}{\phi_\mathrm{M}}\left(\frac{\partial q_1}{\partial t}+\frac{\partial q_2}{\partial t}\right) \equiv -\frac{\rho^\mathrm{B}}{\phi} \frac{\partial q_\mathrm{M}}{\partial t}, \label{MRC eq system cF} \\
 &\frac{\partial c_\mathrm{OH}}{\partial t} = \frac{\rho^\mathrm{B}_\mathrm{M}}{\phi_\mathrm{M}}\frac{\partial q_1}{\partial t}, \label{MRC eq system cOH} \\
 &\frac{\partial q_1}{\partial t} = k_1^\mathrm{a}c_\mathrm{F}(q_1^\mathrm{m}-q_1)-k_1^\mathrm{d}c_\mathrm{OH}q_1, \label{MRC eq system q1} \\
 &\frac{\partial q_2}{\partial t} = k_2^\mathrm{a}c_\mathrm{F}(q_2^\mathrm{m}-q_2)-\kappa_2^\mathrm{d}q_2, \label{MRC eq system q2} 
 \end{align}
 \label{MRC eq system}
\end{subequations}
 \hspace{-1.5mm}where $c_\text{OH}$ is the concentration of hydroxide (mol$/$l), $q_1$ is the 
 moles of the adsorbate PF formed per mass of MRC (mol/g),
$q_2$ is 
the moles of the adsorbate PCa$\leftrightsquigarrow$F formed per mass of MRC (mol/g),
$q_i^\mathrm{m}$ is the maximum attainable value of $q_i$, for $i=1,2$ with $q_\mathrm{M}\defeq q_1+q_2$, and we take $\ast = M$ so that $\rho^\mathrm{B}_\mathrm{M}$ and $\phi_\mathrm{M}$ are the bulk density and porosity, respectively, of the MRC packing. 
We refer to the chemically based model for MRC as defined in Equations (\ref{MRC eq system}) as the CB-MRC model.

 We use the isotherm to fit $k_\mathrm{1}^\mathrm{a}/k_\mathrm{1}^\mathrm{d}$,$k_\mathrm{2}^\mathrm{a}/k_\mathrm{2}^\mathrm{d}$ and $q_\mathrm{2}^\mathrm{m}/q_\mathrm{M}^\mathrm{m}$; subsequently we use the kinetic study to fit $k_\mathrm{1}^\mathrm{a}$ and $k_\mathrm{2}^\mathrm{a}$. 
Table~\ref{MRC new model table} (right columns), shows these parameters. 
The CB-MRC model has a noticeably better fit than Langmuir's model for the isotherm (Figure~\ref{MRC Lang}, left), and most significantly, the kinetic study (Figure~\ref{MRC Lang}, right) has the correct qualitative shape and appears to capture the dominant mechanisms of MRC's removal of F$^-$. As is evident from Table~\ref{MRC new model table}, the MRC model has twice as many fitting parameters as for Langmuir's model, however two-thirds of these values are intrinsic constants and are subsequently used in the column experiments reducing the required number of fitting parameters for the column filter. 

\subsubsection{TMRC} 
 For consistency with the MRC model, we consider the ion exchange between OH$^-$ and F$-$ in the 
 $\text{Al(OH)}_3$ coating of the TMRC. 
 Several ion-exchange reactions occur, both those involving the aluminium and the calcium. As the adsorption capacity of TMRC is approximately ten times greater than that of MRC \cite{chatterjee2018defluoridation}, we neglect any reaction that occurs in TMRC and does not involve aluminium. Further, for a metallic hydroxide with multiple OH$^-$ molecules in an alkaline solution, as is the case in this system, 
the majority of the molecules formed are AlF(OH)$_2$, thus AlF$_2$OH and AlF$_3$ are less 
prevalent \cite{parthasarathy1986study, gayer1958solubility, cowley1948basic}. Appealing to Nie et al. \cite{nie2012enhanced}, we assert that the dominant chemical reaction occurring in TMRC is 
 
 \begin{table}[b]
 \centering
 \footnotesize{
 \begin{tabular}{|c|c|c|c|c|c|c|c|c|}
 \hline 
 \multicolumn{9}{|c|}{\textbf{TMRC isotherm and kinetic parameters (3 s.f.)}} \\ 
 \hline
 \multicolumn{3}{|c|}{\textbf{Extracted (kinetic)}} & \multicolumn{3}{|c|}{\textbf{Optimised: Langmuir}} & \multicolumn{3}{|c|}{\textbf{Optimised: IE-TMRC}}\\
 \hline
 Param. & Value & Units & Param. & Value & Units & Param. & Value & Units \\
 \hline
 $c_\mathrm{F}^\mathrm{i}$ & 5.26$\times 10^{-4}$ & mol/l & \cellcolor[rgb]{0.7922,0.8471, 0.8706}$k_\mathrm{L}^\mathrm{a}/\kappa_\mathrm{L}^\mathrm{d}$ & \cellcolor[rgb]{0.7922,0.8471, 0.8706}$8.47\times10^{3}$ & \cellcolor[rgb]{0.7922,0.8471, 0.8706}l/mol & \cellcolor[rgb]{0.7922,0.8471, 0.8706}$k_\mathrm{T}^\mathrm{a}/k_\mathrm{T}^\mathrm{d}$ & \cellcolor[rgb]{0.7922,0.8471, 0.8706}13.8 & \cellcolor[rgb]{0.7922,0.8471, 0.8706}-- \\
 \hline
 $c_\mathrm{OH}^\mathrm{i}$ & $1.00\times10^{-8}$ & mol/l & \cellcolor[rgb]{0.7922,0.8471, 0.8706}$q_\mathrm{L}^\mathrm{m}$ & \cellcolor[rgb]{0.7922,0.8471, 0.8706}$6.93\times10^{-3}$ & \cellcolor[rgb]{0.7922,0.8471, 0.8706}mol/g & \cellcolor[rgb]{0.7922,0.8471, 0.8706}$q_\mathrm{T}^\mathrm{m}$ & \cellcolor[rgb]{0.7922,0.8471, 0.8706}7.96$\times10^{-3}$& \cellcolor[rgb]{0.7922,0.8471, 0.8706}mol/g\\ 
 \hline
 $c_\mathrm{F}^\mathrm{e}$ & 4.21$\times 10^{-6}$ & mol/l & \cellcolor[rgb]{1,1, 0.8}$k_\mathrm{L}^\mathrm{a}$ & \cellcolor[rgb]{1,1, 0.8}2.31 & \cellcolor[rgb]{1,1, 0.8}l/(mol·s) & \cellcolor[rgb]{1,1, 0.8}$k_\mathrm{T}^\mathrm{a}$ & \cellcolor[rgb]{1,1, 0.8}7.10 & \cellcolor[rgb]{1,1, 0.8}l/(mol·s) \\
 \hline
 \end{tabular}}
 \caption{ 
 Parameters for Langmuir removal (Eqs.~\ref{Lang_MRC}) and for the ion-exchange model for TMRC (IE-TMRC, Eqs.~\ref{TMRC eq system}) as in Table~\ref{MRC new model table} those shaded in grey were determined via fitting the isotherm (Figure~\ref{TMRC pseudo lang}, left) and those shaded in pale yellow were subsequently fitted using the kinetic study (Figure~\ref{TMRC pseudo lang}, right).
 }
 \label{TMRC model table}
\end{table}
\normalsize
\begin{figure}[tb]
 \centering
 \includegraphics[width=.95\textwidth]{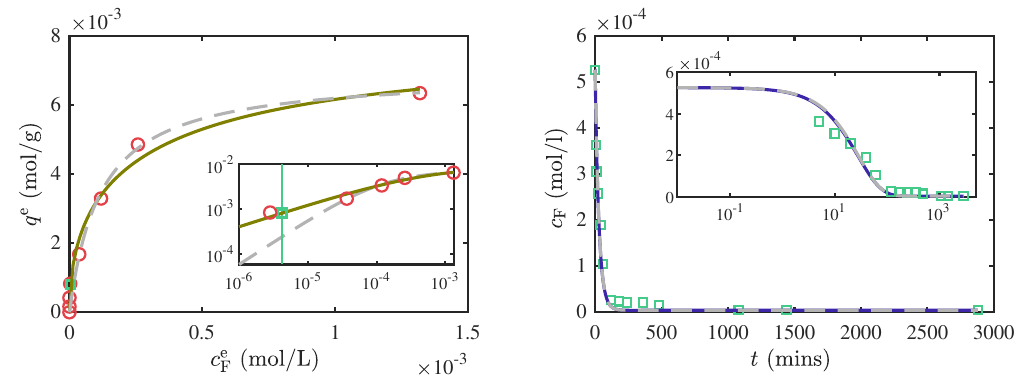}
 \caption{
 Isotherm (left) and kinetic study (right) for TMRC fitted with both the IE-TMRC model (Eqs.~\ref{TMRC eq system}, solid coloured curve) and for reference Langmuir removal (Eqs.~\ref{Lang_MRC}, with $\star =$L, grey dashed curve). All other details are as in Figure~\ref{MRC Lang} but for TMRC in lieu of MRC. 
}
 \label{TMRC pseudo lang}
\end{figure}

\begin{equation}
 \text{Al(OH)}_3+\text{F}^-\stackrel[k_\mathrm{T}^\mathrm{d}]{k_\mathrm{T}^\mathrm{a}}{\rightleftharpoons}\text{AlF(OH)}_2+\text{OH}^-, \label{TMRC reactions 1}
\end{equation}
where $k_1^\mathrm{a}$, $k_1^\mathrm{d}$ are the 
forwards and backwards reaction rates, respectively, of Equation (\ref{TMRC reactions 1}). 
We express Equation~(\ref{TMRC reactions 1}) as a system of differential equations 
\begin{subequations}
 \begin{align}
 &\frac{\partial c_\mathrm{F}}{\partial t} = -\frac{\rho_\mathrm{T}^\mathrm{B}}{\phi_\mathrm{T}}\frac{\partial q_\mathrm{T}}{\partial t}, \label{TMRC eq system cF}\\
 &\frac{\partial c_\mathrm{OH}}{\partial t} = \frac{\rho^\mathrm{B}_\mathrm{T}}{\phi_\mathrm{T}}\frac{\partial q_\mathrm{T}}{\partial t},\label{TMRC eq system cOH} \\ 
 &\frac{\partial q_\mathrm{T}}{\partial t} = k_\mathrm{T}^\mathrm{a}c_\mathrm{F}(q_\mathrm{T}^\mathrm{m}-q_\mathrm{T})-k_\mathrm{T}^\mathrm{d}c_\mathrm{OH}q_\mathrm{T}, 
 \end{align}
 \label{TMRC eq system}
\end{subequations}
\hspace{-1.5mm}where $q_T$ is the number of moles of the adsorbate $\text{AlF(OH)}_2$ formed per mass of TMRC (mol/g), 
$q_\mathrm{T}^\mathrm{m}$ is the maximum attainable value of $q_T$, and $\rho^\mathrm{B}_\mathrm{T}$ and $\phi_\mathrm{T}$ are the bulk density and porosity, respectively, of the TMRC packing. 
We use the isotherm (Figure~\ref{TMRC pseudo lang}, left) to fit two parameters $k_\mathrm{L}^\mathrm{a}/\kappa_\mathrm{L}^\mathrm{d}$ and $q_\mathrm{L}^\mathrm{m}$ for Langmuir's model and $k_\mathrm{T}^\mathrm{a}/k_\mathrm{T}^\mathrm{d}$ and $q_\mathrm{T}^\mathrm{m}$for the ion-exchange model, henceforth referred to as the IE-TMRC model. Subsequently, we use the kinetic study (Figure~\ref{TMRC pseudo lang}, right) to fit $k_\mathrm{L}^\mathrm{a}$ (Langmuir), and $k_\mathrm{T}^\mathrm{a}$ (IE-TMRC). Again, all fittings are via the \verb|GlobalSearch| function in \verb|MATLAB|$^\text{\copyright}$; these values are shown in Table~\ref{TMRC model table}. 

The IE-TMRC model and Langmuir's model are almost indistinguishable in the kinetic study (Figure~\ref{TMRC pseudo lang}, right). 
There is a slight difference between the isotherms of the two TMRC models, however it is not clear from the isotherm (Figure~\ref{TMRC pseudo lang}, left), if one model is significantly better than the other. 
Nie et al. \cite{nie2012enhanced} also find good agreement with the Langmuir isotherm, however they have used a different model (a pseudo-second order reaction) to fit with the adsorption kinetics; this shows an inherent inconsistency which highlights that Langmuir is 
not consistent with the underlying chemistry of this system. 
\normalsize

\subsection{Column filter modelling}
\label{Column}

We consider the steady flow of contaminated fluid through the column filter which we model as a homogeneous porous material. The net flow of the fluid is purely in the $z$-direction (along the length of the filter) with constant Darcy velocity $v$. 
Contaminated fluid enters uniformly at the inlet defined by $z=0$ and exits at the outlet at $z=L$. We control the feed (inlet) concentration, $c_\text{F}^\text{f}$, and measure $c_\text{F}$ at the outlet ($c_\mathrm{F}^\mathrm{out}$) against $t$; this is known as a breakthrough curve. 

 In column filters, advection and shear dispersion (effective diffusivity) are the dominant transport mechanisms for contaminant transport. Thus, we model the transport of $c_\mathrm{F}$ and $c_\mathrm{OH}$ through the filter via advection--diffusion equations, with removal by both MRC and TMRC: 
\begin{subequations}
 \begin{align}
 \label{Combo_cf}
 &\frac{\partial c_\mathrm{F}}{\partial t} = D_\mathrm{F}\frac{\partial^2c_\mathrm{F}}{\partial z^2}-v\frac{\partial c_\mathrm{F}}{\partial z} - \left(\frac{\rho^\mathrm{B}_\mathrm{M}}{\phi_\mathrm{C}}\frac{\partial q_1}{\partial t}+\frac{\rho^\mathrm{B}_\mathrm{M}}{\phi_\mathrm{C}}\frac{\partial q_2}{\partial t}+\frac{\rho^\mathrm{B}_\mathrm{T}}{\phi_\mathrm{C}}\frac{\partial q_\mathrm{T}}{\partial t}\right),\\
 \label{Combo_cOH}&\frac{\partial c_\mathrm{OH}}{\partial t} = D_\mathrm{OH}\frac{\partial^2c_\mathrm{OH}}{\partial z^2}-v\frac{\partial c_\mathrm{OH}}{\partial z} + \left(\frac{\rho^\mathrm{B}_\mathrm{M}}{\phi_\mathrm{C}}\frac{\partial q_1}{\partial t}+\frac{\rho^\mathrm{B}_\mathrm{T}}{\phi_\mathrm{C}}\frac{\partial q_\mathrm{T}}{\partial t}\right), \\ 
 \label{Combo_q1} &\frac{\partial q_1}{\partial t} = k_1^\mathrm{a}c_\mathrm{F}(q_1^\mathrm{m}-q_1)-k_1^\mathrm{d}c_\mathrm{OH}q_1, \\
 \label{Combo_q1}&\frac{\partial q_2}{\partial t} = k_2^\mathrm{a}c_\mathrm{F}(q_2^\mathrm{m}-q_2)-\kappa_2^\mathrm{d}q_2, \\
 \label{Combo_qT}&\frac{\partial q_\mathrm{T}}{\partial t} = k_\mathrm{T}^\mathrm{a}c_\mathrm{F}(q_\mathrm{T}^\mathrm{m}-q_\mathrm{T})-k_\mathrm{T}^\mathrm{d}c_\mathrm{OH}q_\mathrm{T}, 
 \end{align}
 \label{Column eq system}
\end{subequations}
\hspace{-1mm}where $D_\mathrm{F}$ and $D_\mathrm{OH}$ are the effective diffusivities of F$^-$ and OH$^-$, respectively, and where $\phi_\mathrm{C}$ is calculated using a ratio of 40:1 (MRC:TMRC). We refer to the model for MRC-TMRC mixture defined by Equations (\ref{Column eq system}) as the CB-MT model. 

\begin{table}[b]
 \centering
 \begin{tabular}{|c|c||c|c||c|c|}
 \hline
 \multicolumn{6}{|c|}{\textbf{Optimised parameters for column filter (3 s.f.)}} \\ 
 \hline
\textbf{Param.} & \textbf{Langmuir} & \textbf{Param.} & \textbf{CB-MT} & \textbf{Param.} & \textbf{Simplified CB-MT} \\
\hline
$k_\mathrm{L}^\mathrm{a}$ & 0.0192 & $k_1^\mathrm{a}$ & $1.00\times10^{-9}$ & $k_\mathrm{T}^\mathrm{a}$ &0.0564 \\ 
 \hline
 $k_\mathrm{L}^\mathrm{d}$ & $1.66\times10^{-5}$ & $k_2^\mathrm{a}$ & $1.61\times10^{-4}$& \multicolumn{2}{|c|}{}\\ 
 \hline
 $q_\mathrm{L}^\mathrm{m}$ & $7.27\times10^{-4}$ & $k_\mathrm{T}^\mathrm{a}$ & 0.0589&\multicolumn{2}{|c|}{} \\ 
 \hline
 \end{tabular}
 \caption{ \label{tab:HgBC} Fitting parameters for simple Langmuir model (left columns), the CB-MT model (centre columns) and the simplified CB-MT model (right columns). 
 }
\end{table}

\begin{figure}[tb]
 \centering
 \includegraphics[width=.95\textwidth]{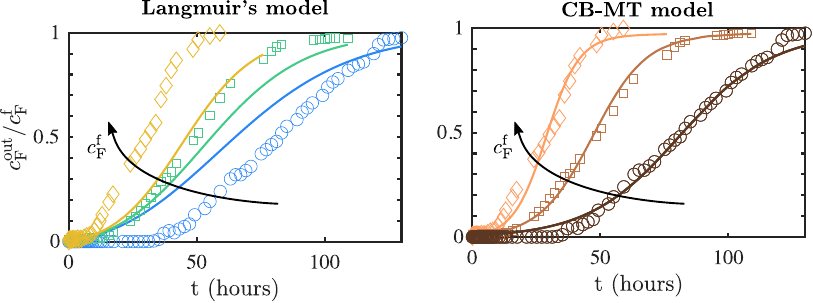}
 \caption{
 The breakthrough curve for a column filter comprising a mixture of MRC and TMRC at a ratio 40:1 (MRC:TMRC), for three different $c_\mathrm{F}^\mathrm{f}$: 5.5mg$/$l (circles), 9.5mg$/$l (squares) and 15.5mg$/$l (diamonds). 
 }
 \label{combo 3 concentration breaktrhough}
\end{figure}

By considering the behaviour of all five quantities ($c_\text{F}, \ c_\text{OH}, q_1^\mathrm{m},\ q_2^\mathrm{m},\ q_\mathrm{T}^\mathrm{m}$) from the CB-MT model at the outlet, in combination with the fact that $k_1^\mathrm{a}$, $k_2^\mathrm{a} \ll k_\mathrm{T}^\mathrm{a}$ (see Table \ref{tab:HgBC}), it is clear that the TMRC dominates the removal despite there being over forty times more MRC (by mass). This is consistent with previous findings; Chatterjee et al. \cite{chatterjee2018defluoridation} find that TMRC has more than ten times the adsorption capacity of MRC. Motivated by this dominance and the inconclusive results of Figure~\ref{TMRC pseudo lang}, we also consider Langmuir removal (an advection--diffusion equation with removal via the sink defined in Equation (\ref{eq: Langmuir})). 
 
 Figure \ref{combo 3 concentration breaktrhough} shows the breakthrough curves for three different $c_\mathrm{F}^\mathrm{f}$: $5\pm0.5$mg$/$l (circles), $10\pm0.5$mg$/$l (squares) and $15\pm0.5$mg$/$l (diamonds). For the purpose of modelling, we take $c_\mathrm{F}^\mathrm{f}\in\{5.5, 9.5, 15.5\}$. For the CB-MT model we use the parameters fitted from the isotherm, as these are intrinsic constants of the physical system, while those determined via the kinetic study do not generally hold. 
For Langmuir's model we cannot use any of the parameters determined from the batch experiments because we have no isotherm curve for the MRC-TMRC mixture. 
However, we do constrain $q_\text{L}^\text{m}$ to be between the values of $q_\text{L}^\text{m}$ obtained via the MRC and TMRC batch experiments --- that is, we take $q_\text{L}^\text{m}\in[7.27\times10^{-4}, 6.93\times10^{-3}]$. 
Table \ref{tab:HgBC} shows the fitting parameters and their values. We fit for a global optimum across all three values of $c_\text{F}^\text{f}$, once again using the \verb|GlobalSearch| function in \verb|MATLAB|$^\text{\copyright}$. 

 The breakthrough curves fit poorly with the optimised Langmuir model (Figure \ref{combo 3 concentration breaktrhough}, left); this strongly indicates that Langmuir is not describing the true behaviour of the system.
 Conversely, the CB-MT model (Figure \ref{combo 3 concentration breaktrhough}, right) fits well with the data for all $c_\text{F}^\text{f}$. 
 Both models have the same number of fitting parameters, hence we conclude that the chemically based model correctly captures the dominant mechanisms of the MRC-TMRC column filter. 

\section{Discussion and Conclusions}
Here, we have presented chemically based models for the adsorbents mineral rich carbon (MRC) and chemically treated mineral rich carbon (TMRC), and their mixture. We have shown how these models compare to classical Langmuir for both batch and column experiments, finding that Langmuir's model fails to produce an acceptable fit with the breakthrough curve despite having three fitting parameters. The chemically based model, also having 3 fitting parameters, has good agreement with the breakthrough curve for all feed concentrations. 
Although TMRC is a significantly better adsorbent than MRC, it is much
more expensive to produce and also due to the fact that the average grain size
is smaller, the presence of too much TMRC leads to clogging in the filter. Thus,
a careful balance between the TMRC and MRC must be reached.
Future work will include comparison with experimental data varying other control parameters such as filter length and flow rate. Also, we will present a simplified model that has just one fitting parameter for the breakthrough curves; we will use this simplified model to predict filter lifespan and efficiency with the aim of design optimisation.

\section*{Acknowledgements}
This publication is part of the research projects PID2020-115023RB-I00 financed by
MCIN/AEI/10.13039/501100011033/ and by “ERDF A way of making Europe” and the CERCA Programme of the Generalitat de Catalunya. The work was also supported by the Spanish State Research Agency, through the Severo Ochoa and Maria de Maeztu Program for Centres and Units of Excellence in R\&D (CEX2020-001084-M). A. V. acknowledges support from the Margarita Salas UPC postdoctoral grants funded by the Spanish Ministry of Universities with European Union funds - NextGenerationEU (UNI/551/2021 UP2021-034).

%

\end{document}